# Words That Stick: Predicting Decision Making and Synonym Engagement Using Cognitive Biases and Computational Linguistics

Dvir, N.,[1] Friedman, E.,[1] Commuri, S.,[1] Yang, F.,[1] Romano, J.[2]
[1] State University of New York at Albany, Albany, New York, NY
[2] Google, New York City, NY

**Corresponding author:**
Nimrod Dvir
Department of Information Systems and Business Analytics
University at Albany, State University of New York
1400 Washington Avenue, Albany, NY 12222
ndvir@albany.edu

## Abstract

This study utilized cognitive psychology and information systems research to predict user engagement and decision making in digital platforms. Applying natural language processing (NLP) techniques and cognitive bias research, we investigated user interactions with synonyms in digital content. Our approach amalgamates four cognitive biases—Representativeness, Ease of use Affect, and Distribution—collectively referred to as the READ model. We evaluated the model's predictive capacity using a large user survey, which revealed that synonyms representative of core concepts that are easy to process, emotionally resonant, and readily available foster increased user engagement. Importantly, our research provides a novel perspective on human–computer interaction, digital practices, and decision-making processes. Our findings underscore the potential of cognitive biases as powerful predictors of user engagement, emphasizing their role in effective digital content design in education, marketing, and beyond.

## 1. Introduction

The rapid rise of digital content consumption has amplified the need for understanding the cognitive processes that drive user engagement. As digital interactions become an integral part of our lives, it is paramount to investigate the factors that influence decision-making processes in these contexts. Cognitive biases, inherent tendencies in human thought that influence our behavior and decisions, offer a rich perspective for understanding such interactions [1]. Recent studies have begun to explore the implications of these biases on user engagement in digital environments, yet there remains a significant gap in the research, particularly in the realm of natural language processing (NLP) and language models, such as generative pre-trained transformers (GPTs) [2, 3, 4].

Artificial intelligence (AI) applications, specifically in the development and improvement of large language models (LLM) like ChatGPT, stand to gain substantial benefits from this line of inquiry [5, 6]. Current LLMs operate based on extensive training on diverse text data with the goal of generating human-like text [7]. Understanding cognitive biases that influence user engagement with digital content could provide novel insights into enhancing the performance and utility of LLMs, ultimately leading to a more intuitive and engaging user experience.

We present a novel model, the READ model, for predicting user engagement with digital content through the lens of four characteristics affected by cognitive biases. We refer to these biases— Representativeness, Ease of use, Affect, and Distribution—as the factors in our READ model. Our research is primarily concerned with the selection and engagement with synonyms in digital content, a crucial factor for AI applications aimed at language generation and understanding [8, 9]. The findings of this study contribute to the broader discourse on human-computer interaction (HCI) and the role of cognitive factors in shaping digital experiences.

Our research has the potential to provide important insights into how LLMs can be trained to better align with human cognitive processes. The READ model's predictive accuracy in estimating synonym selection paves the way for development of more personalized and engaging AI applications. This work is an attempt to bridge the gap between cognitive neuroscience and information systems (IS) research by offering a new lens through which we can understand and enhance human–technology experiences.

## 2. Literature Review

Our research resides at the intersection of cognitive linguistics, NLP, and user engagement. These fields have been extensively studied, albeit independently, giving us a rich pool of literature from which to draw.

### 2.1 Cognitive Biases and Linguistics

The concept of cognitive biases was first introduced by Tversky and Kahneman [10], who identified universal and systemic errors in human decision making and judgment. These biases are inherent in cognition, often influencing how we perceive and interpret information. Although cognitive biases have been predominantly studied in the realm of psychology and decision sciences, recognition of their impact on linguistics has emerged, and research into their impact has been recognized as crucial [11, 12].

Kahneman emphasized that language comprehension and production are significantly influenced by these biases [1]. In accordance, research indicates that cognitive biases can considerably impact the choice and perception of words, particularly synonyms [3, 4, 13–15]. Thus, the understanding of cognitive biases in the context of linguistic preferences and user engagement holds considerable potential for research and application.

### 2.2 Natural Language Processing and Synonyms

NLP, a subfield of artificial intelligence, studies the interaction between computers and human language [16, 17]. Understanding of synonyms is a critical aspect of NLP, as evidenced by the substantial research into synonym recognition, generation, and substitution [4, 13, 18].

While synonyms are semantically equivalent, they often carry different connotations, nuances, and cognitive implications [13, 19]. This characteristic makes knowledge of synonym choice a rich and potent tool for predicting user engagement and response. Miller's development of WordNet, a lexical database, has greatly facilitated synonym understanding in NLP, and provided a robust platform for our research [20].

### 2.3 User Engagement

User engagement has evolved as a crucial aspect of digital content research, particularly research into the factors that initiate and maintain attention [21, 22]. Prior research indicates that the phrasing used in content, specifically word choice, significantly influences user engagement [9].

## 3. Conceptual Model

Our conceptual model fuses insights from the fields of cognitive linguistics, NLP, and user engagement to harness the potential of cognitive biases in predicting user engagement with synonym usage. Our research innovatively combines these domains to propose a unique approach that leverages cognitive biases and NLP techniques to predict user engagement. Rooted in four cognitive biases, our READ model provides a theoretical framework with which to evaluate and predict user engagement with synonyms in digital content.

To the best of our knowledge, this is the first study to examine user engagement from this perspective and the READ model is the first model based on the findings thereof, marking a significant advancement in the field.

### 3.1 Representativeness

The representativeness bias, also known as the representativeness heuristic, is a common cognitive shortcut used for making judgments of probability by estimating the probability that a novel occurrence resembles (i.e., is representative of) an exemplary occurrence [12, 13, 23].

Translating this bias to our model, we argue that users prefer synonyms that best represent an underlying concept or idea, and hence engage more frequently with content using them. Based on this argument, we propose Hypothesis 1A:

*Hypothesis 1A: Synonyms that are more representative of an underlying concept or idea will have a higher selection rate.*

### 3.2 Ease of use

Ease of use, also referred to as processing fluency, refers to the ease with which information is processed. Research shows that individuals prefer, and even find more believable, information that is simple to understand and use [24, 25, 26]. In the context of our model, we argue that synonyms that are easy to process will garner more attention and engagement from users. Based on this argument, we propose Hypothesis 1B:

*Hypothesis 1B: Synonyms that are easier to process will have a higher selection rate.*

### 3.3 Affect

Affect, specifically affect-biased attention, refers to a bias resulting from automatic prioritization of emotionally or motivationally salient stimuli [14, 27, 28, 29]. Applied to our model, we argue that synonyms carrying more emotional weight will capture user attention more effectively, driving higher engagement levels. Based on this argument, we propose Hypothesis 1C:

*Hypothesis 1C: Synonyms with higher emotional resonance will have a higher selection rate.*

### 3.4 Distribution

The distribution bias, also referred to as the availability bias, refers to relying on readily available information for making rapid decisions and judgments [10, 26, 30]. In our model, we argue that synonyms easily retrievable from memory will generate higher user engagement. Based on this argument, we propose Hypothesis 1D:

*Hypothesis 1D: Synonyms that are more readily available will have a higher selection rate.*

Collectively, our hypotheses contribute to our overarching hypotheses:

*Hypothesis 2: Synonyms with higher selection rates will have significantly higher levels of representativeness, ease of use, affect, and distribution.*

*Hypothesis 3: Synonym selection rate can be predicted based on level of representativeness, ease of use, affect, and distribution.*

## 4. Methodology

### 4.1 Data Collection

We chose 50 pairs of synonyms for our study from WordNet, a robust English lexical database [20]. WordNet groups words into synsets, each representing a unique meaning. We selected 50 common English words and their corresponding synonyms from various synsets, resulting in 100 words for analysis. We calculated the READ values for each word based on a substantial user survey.

### 4.2 Variable Measurement

Using a custom Python software program incorporating knowledge of computational linguistics and NLP techniques, we evaluated textual features to measure the READ factors.

**4.2.1** **Representativeness.** We measured the representativeness of synonyms via semantic similarity measures obtained from WordNet. We determined each word's number of meanings, synonyms, hypernyms (broader terms), and hyponyms (more specific terms).

**4.2.2** **Ease of use.** We measured ease of use based on each synonym's word length and syllable count.

**4.2.3** **Affect.** We measured affect using SentiWordNet 3.0, a WordNet-based lexical resource, to assign sentiment scores— Positivity, Negativity, and Emotionality scores—to each synonym that reflected their emotional resonance [31].

**4.2.4** **Distribution.** We determined synonym distribution by determining their frequency of use utilizing the findings of Kuperman, Stadthagen-Gonzalez, and Brysbaert [32] and WordFreq, and using a Python library that provided data from multiple large text corpora to approximate word usage in natural language [33].

## 4.3 Model Formulation

We integrated our variable measurements to formulate a multivariate predictive model estimating synonym selection rates.

## 4.4 Validation

We validated our model using a comprehensive user survey implemented through QualtricsXM, a cloud-based platform. The survey asked undergraduate students from a US research university select their preferred synonyms from groups of synsets. We collected 80,500 observations in total, carefully controlling for unique participants and participant characteristics to mitigate selection and allocation biases. Pearson chi-square analysis confirmed no significant compositional differences between demographic groups and word samples.

## 5. Results

All data collected in this study, including the 50 synset pairs and specific words used, their selection rates, and READ features is available publicly for inspection and replication (https://dataverse.harvard.edu/dataset.xhtml?persistentId=doi:10.7910/DVN/J5LTYE) [34]. We collected 80,500 responses, thereby obtaining 805 responses per word. The mean selection rate for specific words was 0.22 (SD = 0.017).

## 5.1 Correlation between selection rates and cognitive biases

**5.1.1** **Representativeness.** The Pearson correlation coefficients for the relationship between a word's selection rate and number of definitions ($r = 0.207$, $p < 0.001$), synonyms ($r = 0.200$, $p < 0.001$), hypernyms ($r = 0.269$, $p < 0.001$), and hyponyms ($r = 0.155$, $p < 0.001$) were significantly positive. The analysis of our user survey data indicated statistically significant positive correlations between a synonym's representativeness and its selection rate, This significantly positive relationship underscores the pivotal role of representativeness in user engagement, supporting Hypothesis 1A.

**5.1.2** **Ease of use.** The Pearson correlation coefficients for the relationship between a word's selection rate and its length ($r = -0.384$, $p < 0.05$) and number of syllables ($r = -0.410$, $p < 0.05$) were statistically significant. As shorter words and words with fewer syllables are considered simpler and easier to process, there is a significantly positive correlation between the processing fluency (ease of understanding) of a synonym and its selection rate, supporting Hypothesis 1B.

**5.1.3** **Affect.** The Pearson correlation coefficients between the selection rate of words and their positivity ($r = 0.137$, $p < 0.01$), negativity ($r = 0.117$, $p < 0.01$), and overall emotionality ($r = 0.184$, $p < 0.01$) values were statistically significant, affirming the role of affect-biased attention in user engagement with synonyms. Specifically, synonyms with higher emotional resonance, be it positive or negative, garner higher user engagement, supporting Hypothesis 1C.

**5.1.4 Distribution.** The Pearson correlation coefficient between the selection rate of synonyms and their frequency of was statistically significant ($r = 0.018$, $p < 0.001$). This finding indicates that synonyms that are more frequently used, and thus more readily available to users, are selected at a higher rate, supporting Hypothesis 1D

## 5.2 READ Scores Among Synonyms

To explore the differences in READ measures between highly selected and less selected synonyms, we performed independent sample t-testing for all 50 pairs of synonyms. We found that 16 pairs had notable differences in their mean selection rates, marked by a t-value of ±1.961 and a significant p-value of less than 0.05G. When we conducted binomial testing to discern whether the observed proportion of significant results deviated from what might be expected by chance, we found that the proportion of significant results in our study (0.16, derived from 16 out of 100) was markedly higher than chance predictions (0.05).

Delving deeper into the R.E.A.D measures, we conducted additional independent sample t-testing between synonyms with significantly higher selection rates and their counterparts with lower rates, with equal variance not assumed. The results revealed significant differences across multiple measures.

The number of definitions, a factor in a word's representativeness, had a high t-value of 102.016 and a highly significant p-value of less than 0.001 with a large effect size (Cohen's d = 2.573. We found that the number of synonyms, another factor in representativeness, had a high t-value of 92.240, a p-value of less than 0.001, and a large effect size (Cohen's d = 5.460).

Hypernym and hyponym counts, additional factors in representativeness, also had significant differences with t-values of 100.430 and 40.524, p-values of less than 0.001, and effect sizes of 1.766 and 9.977, respectively.

Regarding ease of use, both syllable count and word length had negative t-values of -121.140 and -140.503, p-values of less than 0.001, and large effect sizes of 0.820 and 1.615, respectively, indicating that synonyms with shorter lengths and fewer syllables had higher selection rates because they were easier to use and process.

Exploring the role of affect, the maximum negativity sentiment (NegMax) had a significant difference with a t-value of 8.904, a p-value of less than 0.001, and a small effect size (Cohen's d = 0.298743), while the maximum positivity sentiment (PosMax) had a negative t-value of -7.340, a p-value of less than 0.001, and a small effect size (Cohen's d = 0.241575). Interestingly, this indicates that synonyms with higher selection rates are associated with higher negative sentiment, while those with lower selection rates are associated with higher positive sentiment.

Lastly, the word frequency, a measure of distribution, had a t-value of 94.004, a p-value of less than 0.001, and a negligible effect size (Cohen's d = 0.0000645), indicating that synonyms with higher selection rates appear more frequently in the language.

In alignment with Hypothesis 2, these results suggest that synonyms with significantly higher selection rates are more representative, easier to use, more negatively emotional, and more readily available. This set of findings underscores the impact of cognitive biases on user engagement with digital content.

## 5.3 Hypothesis 3

Our research focused on examining the variation in selection rates between pairs of synonyms and its correlation with differences in the four cognitive biases, namely representativeness, ease-of-use, affect, and distribution (READ), that form the basis of our model. We calculated this variation by subtracting the selection rate of one synonym from the other within each pair. This method allowed us to quantify the difference in user engagement between the two synonyms in each set.

In terms of potential predictors, we adopted a similar approach for various features across the READ measures. For each synonym pair, we computed the difference in each feature by subtracting the value of one synonym from that of the other. This process helped us quantify the disparities within each READ measure between the two synonyms in each pair.

This approach allowed us to use the differential values across the READ measures as our predictor variables in the model. The dependent variable was the variation in the selection rate among the synonym pairs. The aim of our multivariate linear regression analysis was to predict this variation based on the differences in READ variables.

The results of independent sample t-testing of the 50 pairs of synonyms revealed that 16 pairs showed significant differences in mean selection rates ($t = \pm1.961$, $p < 0.05$). We corroborated this observation with the results of binomial testing ($p = 0.000037$), indicating that the proportion of significant results in our study was significantly higher than what might be expected from chance alone.

The multivariate linear regression model exhibited significant results ($F[1, 98] = 13.096$, $p < 0.001$), suggesting a good fit of the model to the data. The model accounted for approximately 54.5% of the variance in the dependent variable, as indicated by an R-squared value of 0.545. Regarding case-wise diagnostics, our model accurately predicted the direction of the difference between synonyms in 88% of cases, signifying that it successfully anticipated which synonyms would be more engaging in 88% of the word pairs or in 44 out of 50 instances.

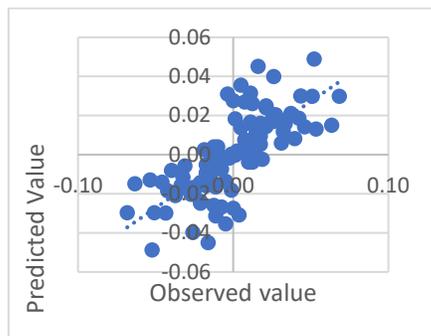

*Figure 1. Predicted versus observed values*

Our model's high predictive accuracy in estimating the selection rates of synonyms based on their READ values strongly support Hypothesis 3 by strongly indicating that cognitive biases are indeed powerful predictors of user engagement with synonyms in digital content.

## 6. Discussion

Our findings underscore the significant predictive capacity of cognitive biases in synonym selection and user engagement with digital content. Our findings pave the way for future investigations in this area, offering valuable guidance for enhancing communication across a range of fields.

### 6.1 Cognitive Bias Measures as Predictors of Synonym Selection

Our findings reveal a strong correlation between a word's representativeness (i.e., number of definitions, synonyms, hyponyms, and hypernyms), ease of use (i.e., word length and number of syllables), affect (i.e., positive, negative, or neutral sentiment), and distribution (i.e., frequency). Our findings substantiate the argument that synonyms more representative of a core concept have higher selection rates, suggesting that using synonyms of higher representativeness may augment engagement in digital content. Similarly, our findings affirm that synonyms that are simpler and easier to process have higher selection rates, reinforcing processing fluency's vital role as a predictor of user engagement. Likewise, our findings that emotionally charged words tend to draw more attention, thereby boosting user engagement and affirming the role of affect in synonym selection.

Our multivariate linear regression model successfully predicted the synonym with a higher selection rate in 88% of synsets, providing robust support for our conceptual model and affirming the influential role of cognitive biases in shaping user engagement with synonyms. Our findings provide robust evidence supporting the impact of distribution on synonym selection and subsequent user engagement.

### 6.2 Implications for Messaging and Communication Strategies

Our findings indicate that understanding and exploiting cognitive biases can facilitate the creation of more compelling messaging and communication strategies. In the context of education, these insights can aid in developing materials that foster superior comprehension and retention. Using words that are easy to process, emotionally relatable, representative, and readily available can assist in creating learning materials that are appealing and engaging. In marketing, word choice can considerably sway the success of a campaign or strategy. Marketers can select synonyms in line with the cognitive biases identified in our study to devise messages that capture attention, foster emotional bonds, and ultimately stimulate customer engagement and action.

### 6.3 Limitations and Future Research

Despite offering significant insights, our study has several limitations that may limit the generalizability of our findings. The first pertains to the selection of synonyms, which was restricted to common English words. To overcome this limitation, future research should aim to incorporate less commonly used words or extend this line of investigation to other languages. The second limitation was reliance on a single measure for each bias, which may not have fully encapsulated the complexity of each cognitive bias. Subsequent studies should integrate multiple measures for each bias to allow a more comprehensive analysis. A third limitation was the investigation of only a small number of factors in our model. Although our model demonstrated high predictive accuracy, the potential influence of other factors, such as cultural, social, and contextual factors, on synonym selection and user engagement should be investigated. A final limitation was that our quantitative analysis might not have captured the nuanced motivations underlying synonym selection. Future research should use qualitative methods, such as in-depth interviews or focus groups, to gain a more profound understanding of word selection.

## 7. Conclusion

With our model we aspires to provide a comprehensive, cognitive bias-based framework for predicting user engagement with synonyms in digital content. Our research uncovered the role of cognitive biases in predicting user engagement with synonym usage in digital content. We proposed a novel conceptual model integrating the representativeness, ease of use, affect, and distribution biases. The model provides a comprehensive, cognitive-based framework to predict user engagement with digital content based on our research into the predictive power of these biases in synonym selection.

The evidence from our study confirmed our hypotheses that synonyms that are more representative of an underlying concept, easier to process, more emotionally resonant, and more available to users will garner more engagement. Our model was successful in predicting synonym selection rates based on these cognitive biases.

These findings contribute to the existing body of knowledge in NLP, cognitive linguistics, and user engagement by presenting new insights into the cognitive aspects that shape user interactions with language. It widens our understanding of how cognitive biases affect language use and paves the way for further investigations in this exciting cross-disciplinary field.

The practical implications of our research are far reaching. In the field of education, this understanding of word choice can enhance the design of learning materials, making them more engaging and accessible to students. in the field of marketing, choosing the more engaging synonyms based on our model could lead to more successful and resonant messaging. More generally, the ability to predict user engagement with different synonyms could improve communication strategies across numerous domains.

As we look forward, there are several avenues for future research. These include expanding the scope of the study to the study of other languages, the use of multiple measures to evaluate the effect of biases, consideration of other factors influencing synonym selection, and employment of qualitative research methods to provide richer insights.

Our study underscores the potential of cognitive biases as powerful predictors of user engagement with synonyms in digital content. By leveraging these insights, we can create more engaging and effective communication, enhancing our ability to connect, inform, and influence through language. The journey ahead promises to be a fascinating exploration of the intersection of cognitive science, linguistics, and digital engagement.